\documentclass[12pt]{article}

\usepackage{graphicx}
\usepackage{longtable}

\begin{document}

\title{Soliton with a Pion Field in the Global Color Symmetry Model}

\author{{Bin Wang$^{1}$, Hui-chao Song$^{1,}$\footnote{present address:
Department of Physics, Ohio State University, Columbus, OH43210,
USA}, Lei Chang$^{1}$, Huan Chen$^{1}$, and Yu-xin
Liu$^{1,2,3,}$\footnote{Corresponding author} }    \\
\normalsize{$^{1}$ Department of Physics, Peking University,
Beijing 100871, China} \\
\normalsize{$^{2}$ The Key Laboratory of Heavy Ion Physics,
Ministry of Education, Beijing 100871, China } \\
\normalsize{$^{3}$ Center of Theoretical Nuclear Physics, National
Laboratory of Heavy Ion Accelerator,}\\
\normalsize{Lanzhou 730000, China} }

\date{\today}

\maketitle

\begin{abstract}
We calculate the property of the global color symmetry model
soliton with the pion field being included explicitly. The
calculated results indicate that the pion field provides a strong
attraction so that the eigen-energy of a quark and the mass of a
soliton reduce drastically, in contrast to those with only the
sigma field.
\end{abstract}

{\bf PACS Numbers:} 12.39.-x, 03.75.Lm, 12.40.Yx, 11.10.Lm


\newpage

\parindent=20pt

\section{Introduction}
Quantum Chromodynamics (QCD) has been accepted as the most
successful theory to describe the strong interaction. However, the
low energy property, such as the property of baryon, is hard to be
described with the precise QCD. Lattice QCD is then developed to
calculate the property of baryons from the full QCD lagrangian
(see for example Ref.\cite{Lepage}). Meanwhile, on the side of
continuous field theoretical method, there are lots of
phenomenological models to try to give the picture of the baryons.
Soliton models, including the two most successful models, Skyrme
model(or chiral soliton model, see for example Ref.\cite{Skyrme})
and the chiral quark soliton model(or Nanbu--Jona-Lasinio (NJL)
Model, see for example Ref.\cite{NJL}), are one kind of these
models to give a vivid picture of baryons. However, the Skyrme
model and the chiral quark soliton model give two contrary
pictures. In Skyrme model, baryons are described as a soliton of
mesons with no valence quark degrees of freedom. In contrast to
Skyrme model, baryons are described as valence quarks bounded in
the meson soliton background in the chiral quark soliton model.
However, only local interactions are taken into account in NJL
model.

To search for the solid QCD foundation of the soliton models and
implement the nature of non-locality of the quarks in mesons, the
global color symmetry model (GCM) has been developed since the
middle of 1980's\cite{GCM85,GCM87,GCM88,formula-Fpi,LLZZ98}. In
the GCM, the gauge symmetry was discarded and then QCD is reduced
to a finite-range current-current interaction theory. One can get
a quark-meson interaction model or a quark-diquark interaction
model after bosonization\cite{GCM87} of the current-current
interaction theory. Since there are many kinds of meson fields to
interact with quarks and it is hard to deal with all of them, one
usually takes the low energy effective mode, the Goldstone mode
pion and sigma, and discards the others. Then light baryons are
modelled as solitons in the chiral meson background with Goldstone
mesons\cite{GCMsoliton1,GCMsoliton2,GCMsoliton3}. With the GCM
soliton, one can also discuss the dynamical confinement and, in
particular, the effect of the non-locality. Furthermore, the
dependence of baryon properties on hadron matter medium has also
been studied in the
GCM\cite{FT97,BR968,BRS98,MRS98,BGP99,MT9914,Liu0135}. However, in
the practical calculations, one usually takes the chiral circle
constraint, i.e., solves the coupled equations of quark and
$\sigma$-meson field explicitly, but discards the pion fields.
Then in this paper, we extend the GCM soliton to beyond the chiral
circle constraint, i.e., deal with the scalar field $\sigma$ and
the pseudoscalar field $\pi$ simultaneously. We can recognize that
there is much difference between the results with and beyond the
chiral circle constraint.

Below, we will give a concise description of the GCM model and of
the soliton in GCM model in section II (an extensive review of GCM
model can be found in series of papers\cite{GCMreview1,
GCMreview2}). In section III, we describe the method of the
numerical calculation and give our numerical results and compare
the results with those with the chiral circle constraint. Finally,
we summarize our work and give a brief remark in section IV.

\section{Brief Review of GCM}
The global color symmetry model (GCM) is constructed by the the
generation functional in Euclidean space as\cite{GCM85,GCM87}
\begin{equation}\label{GCMZ}
Z \! = \! \int \! \mathcal{D}\bar{q}(x)\mathcal{D}q(y) \exp[-\!\!
\int\!\! d^4x \bar{q}(x)\!{\not{\partial}q(x)} - \frac{g^2}{2}\!\!
\int\!\!\!\int \!\! d^4x d^4y \bar{q}(x)\gamma_{\mu}
\frac{\lambda^a}{2} q(x)D(x\!- \! y)\bar{q}(y) \gamma_{\mu}
\frac{\lambda^a}{2}q(y)] \, .
\end{equation}
This functional is invariant under global color $SU(3)$
transformation rather than the gauge color $SU(3)$ transformation.
$D(x-y)$ is the effective gluon propagator, which is a
parameterized function to model the low energy property and
dynamics such as those of hadrons. It has been shown that the
infrared enhancement of $D(x-y)$ is closely related to the chiral
symmetry breaking (see for example Refs.\cite{RW94,Pen04}). Such a
character should be embedded in the model.

It is convenient to calculate the Grassmann integration after
bosonization\cite{GCM85}. Firstly, one can reorder the quark
fields through the Fierz transformation and get
\begin{equation}
Z=\int\mathcal{D}\bar{q}\mathcal{D}q \exp \left[-\int
\bar{q}\not{\partial}q-\frac{g^2}{2}\int\int\bar{q}(x)
\frac{M^\theta}{2}q(y)D(x-y)\bar{q}(y)\frac{M^\theta}{2}q(x)
\right] \, ,
\end{equation}
where $M^\theta = K^a \otimes C^b \otimes F^c$, with $\{K^a\} =
\{I,i\gamma_5,\frac{i}{\sqrt{2}}\gamma_\mu,
\frac{i}{\sqrt{2}}\gamma_5\gamma_\mu\}$, $\{ C^b \}=\{
\frac{4}{\sqrt{3}}I,\frac{i}{\sqrt{3}}\lambda\}$, $\{F^c \}=\{
\frac{I}{\sqrt{2}},\frac{\vec{\tau}}{\sqrt{2}} \} $, are the
direct product of spin, color $SU(3)$ and flavor $SU(2)$ matrices
resulting from Fierz transformation. Then, one introduces bi-local
Bose fields $B^\theta(x,y)$ which transform like
$\bar{q}(x)\frac{M^\theta}{2}q(y)$. The bilocal fields have the
quantum numbers as those of mesons, so the fluctuations on vacuum
can be identified as mesons. One obtains then
\begin{equation}
 Z=\int\mathcal{D} B^\theta(x,y) e^{-S[B^\theta(x,y)]} \, ,
\end{equation}
with an action given as
\begin{equation}
S[B^\theta(x,y)]\! = \! \int \! d^4x d^4y \bar{q}(x) \left[\gamma
\cdot \partial\delta(x\!- \! y) + \frac{M^\theta}{2}B^\theta(x,y)
\right]q(y) + \int d^4x d^4y \frac{B^\theta(x,y)B^\theta(y,x)} {2g^2
D(x-y)}\, .
\end{equation}
Introducing the quark chemical potential through the transformation
\begin{equation}\label{}
  q(x) \rightarrow e^{\mu x_4}q(x)
\end{equation}
and integrating the quark fields, one gets the action
\begin{equation}
S[\mu,B^\theta]=-Tr\ln G^{-1}[\mu,B^\theta]+\int d^4x d^4y
\frac{B^\theta(x,y)B^\theta(y,x)}{2g^2 D(x-y)}\, ,
\end{equation}
where the inverse of the quark propagator can be written as
\begin{equation}
 \begin{array}{rcl}
  G^{-1}(\mu;x,y) &=& e^{\mu x_4}G^{-1}(x,y)e^{-\mu y_4}  \\
                  &=& (\gamma\cdot\partial-\mu\gamma_4)\delta(x-y)
                      + e^{\mu x_4}\frac{M^\theta}{2}B^\theta(x,y)
                        e^{-\mu y_4} \, .
 \end{array}
\end{equation}
Generally, the bilocal fields can be written as
\begin{equation}\label{form}
B^\theta(x,y)=B_0^\theta(x,y)+\sum\limits_i \Gamma_i^\theta(x,y)
\phi_i^\theta(\frac{x+y}{2}) \, ,
\end{equation}
where the first term is the translation invariant vacuum
configuration of the bilocal fields. The second term is the
fluctuation over the vacuum which can be identified as effective
meson fields. $\Gamma_i^\theta$ is the form factor of the meson
fields, which we take as the form of $B_0^{\theta}$ with
appropriate Lorentz structure since the excitation of the inner
degrees of freedom needs much energy and can be frozen in low
energy range\cite{LLZZ98}. There are many meson fields, including
scalar, pseudoscalar, vector, axial vector and tenser field. In
the lowest order, we can only consider the Goldstone mode,
$\phi_i^\theta=\{ \sigma,\vec{\pi} \}$, which was thought of as
the most important low energy degrees of freedom.

The vacuum configuration can be determined by the saddle point
condition $\frac{\delta S}{\delta B_0^\theta}=0$, which induces a
translation invariant quark self-energy $\Sigma(x-y) =
\frac{M^\theta}{2}B_0^{\theta}(x,y) =
\frac{M^\theta}{2}B_0^\theta(x-y) $. The equation of $\Sigma(x-y)$
in momentum space is, in fact, a truncated Dyson-Schwinger
equation, which reads
\begin{eqnarray}\label{DSE}
\Sigma(p) &=& i\gamma\cdot p [A(p^2)-1]+B(p^2)  \nonumber \\
          &=& g^2\int \frac{d^4q}{(2\pi)^4} D(p-q) \frac{\lambda^a}{2}
          \gamma_\mu \frac{1}{i\gamma\cdot q + \Sigma(q)}
          \gamma_\mu \frac{\lambda^a}{2} \, ,
\end{eqnarray}
where $D(p)$ is the Fourier transformation of $D(x)$.

It is evident that the quark meson coupling is in Yukawa form
while the meson self-interaction is nontrivial. The dynamical
chiral symmetry breaking is due to $B(p^2)\neq 0$, which causes
quark a dynamical mass $B(p^2)/A(p^2)$. One can model the low
energy property through a certain form of $A(p^2)$ and $B(p^2)$
determined by solving Eq.(\ref{DSE}) with an effective gluon
propagator or derived from some other models such as instanton
model, or lattice stimulation.

In the chiral limit, the bi-local fields can be written as
\begin{equation}
\frac{M^\theta}{2}[B^\theta
(x,y)-B_0^\theta(x,y)]=\frac{B(r)}{f_\pi} \chi(R) e^{i\gamma_5
\tau\cdot\phi(R)/f_\pi} ,
\end{equation}
where $r=x-y$ and $R=\frac{x+y}{2}$ can be considered as the
relative coordinate and centroid coordinate of a quark and a
anti-quark in a meson, respectively.

Expanding the action to leading order in derivatives of
$\sigma=\chi \cos[{\phi} / {f_\pi}]$ and $\vec{\pi}=\hat{\phi}\chi
\sin[{\phi}/{f_\pi}]$ at the vacuum $\sigma = f_\pi,\vec{\pi}=0$,
one gets
\begin{equation}
S[\sigma,\vec{\pi}]-S[f_\pi,0]=\int d^4R
\{\frac{1}{2}(\partial_\mu \sigma)^2+\frac{1}{2}(\partial_\mu
\vec{\pi})^2+U(\chi^2(R))  \} ,
\end{equation}
where $U(\chi^2)$ is the effective potential of mesons with
$\chi^2 = \sigma ^2 + \vec{\pi}^2 $ and can be given as
\begin{equation}
U(\chi^2)=-12\int \frac{d^4q}{(2\pi)^4} \left\{ \ln\left[
\frac{q^2A^2(q^2)+B^2(q^2)(\chi/f_\pi)^2}{q^2A^2(q^2)+B^2(q^2)}
\right] - \frac{B^2(q^2)[(\chi/f_\pi)^2-1]}{q^2A^2(q^2)+B^2(q^2)}
\right\} ,
\end{equation}
and $f_\pi$ is the pion decay constant given by
\begin{equation}\label{fpi}
f_\pi=12\int \frac{d^4q}{(2\pi)^4} \left[ \frac{B^2
A^2}{[q^2A^2+B^2]^2}-\frac{\frac{1}{2}q^2[(B^\prime)^2 +
BB^{\prime\prime}]+BB^\prime}{q^2A^2+B^2} \right] .
\end{equation}
It should be mentioned the above expression of $f_\pi$ is only
correct when the derivatives of $A(p^2)$ are neglected. However,
this approximation can be fairly right according to the
characteristic of the effective gluon propagator in the present
calculation. The full formula to evaluate the $f_\pi$ can be found
in \cite{formula-Fpi}.

The meson masses can be obtained by differentiating the potential
$U$ twice with respect to $\sigma$ and $\pi$. It can be found that
the pion mass is zero and the $\sigma$ mass is finite at the chiral
limit.

Originally, baryons are regarded as solitons with bag
constant\cite{GCM85}
$$ {\cal{B}} = U(0) - U( f_{\pi} ) \, , $$
where $U(0)$ is the effective potential of mesons in the soliton
with assumption $\sigma = 0 $, $\vec{\pi} = \vec{0}$ and $U(
f_{\pi})$ the one of vacuum with $\sigma = f_{\pi}$, $\vec{\pi} =
\vec{0}$. Even though some baryon properties and their dependence on
baryon density have been obtained in such a model (see for example
Refs.\cite{GCM85,Liu0135}), the advantage of the quark-quark
interaction in the GCM has not yet been represented explicitly. Then
a much more elaborate soliton model was developed in
1990's\cite{GCMsoliton2}. As in Ref.\cite{GCMsoliton2}, the meson
fields in the soliton are written explicitly as
\begin{equation}
\frac{M^\theta}{2} [B^\theta (x,y)-B_0^\theta(x,y)] = B(x-y)\Big[
\sigma(\frac{x+y}{2})+i\gamma_5
\vec{\tau}\cdot\vec{\pi}(\frac{x+y}{2}) \Big] \, .
\end{equation}
After introducing a chemical potential of quarks as the above, one
expresses the action-difference between the vacuum state and a
system with nonzero chemical potential as
\begin{equation}
S[\mu,\sigma,\vec\pi]=Tr\ln G^{-1}[\mu \! = \! 0,\sigma,\vec\pi] -
Tr\ln G^{-1}[\mu,\sigma,\vec\pi] \! - \! \int d^4x [ \frac{1}{2}
(\partial_x \sigma)^2 + \frac{1}{2}(\partial_x \vec\pi)^2 +
U(\chi^2) ] \, .
\end{equation}
where $\chi^2=\sigma^2+\vec{\pi}^2$. The inverse of quark
propagator at $\mu=0$ is given as
\begin{equation}
G^{-1}(\mu=0;x,y)=\gamma\cdot\partial_x \delta(x-y)
+\frac{1}{f_\pi}B(x-y)[\sigma(\frac{x+y}{2}) + i
\gamma_5\vec\pi(\frac{x+y}{2})] \, ,
\end{equation}
and that at finite $\mu$ can be written as
\begin{equation} G^{-1}(\mu;x,y)=e^{\mu x_4}G^{-1}(\mu=0;x,y) e^{-\mu
y_4} .
\end{equation}
In the vacuum, the fields $\sigma$ is $f_\pi$ and $\vec\pi$ is
zero determined by the saddle point condition of $S[\sigma,\pi]$.
When there are valence quarks, the meson fields will vary with the
spatial coordinate to show the shape of the bound system, the
hadron.

One can define an effective potential through the Legendre
transformation. With the static meson field, the effective
potential is proportional to the energy functional. At the Hartree
level as in Ref.\cite{GCMsoliton1}, the energy functional of the
static meson fields at a set of fixed quark occupation number $n$
is
\begin{equation}
  E[n,\sigma,\vec\pi]=E_q[n,\sigma,\vec\pi]+E_m[n,\sigma,\vec\pi] ,
\end{equation}
where $E_q$ is the valence quark's contribution,
\begin{equation}\label{Eq}
E_q[n,\sigma,\vec\pi]\left[ -\int d^4x \right] = Tr\ln
G^{-1}[\mu,\chi]-Tr\ln G^{-1}[0,\chi]-\mu n  \, ,
\end{equation}
and $E_{m}$ the meson field's contribution
\begin{equation}
E_m[n,\sigma,\vec\pi]=\int d^3x \left[ \frac{1}{2}(\nabla
\sigma)^2+\frac{1}{2}(\nabla \vec\pi)^2+U(\chi) \right] \, .
\end{equation}
The quark chemical potential $\mu$ is a functional of the meson
fields $\sigma$ and $\pi$ and the quark occupation number $n$,
which can be deduced from the relation $n=\partial Tr\ln
G^{-1}[\mu;\sigma,\vec\pi] / \partial \mu$.

It is convenient to calculate the above trace of the quark
propagator in Eq.(\ref{Eq}) at certain bases, and one can get the
bases through the spectral decomposition of the quark propagator.
With the static meson fields, $G^{-1}(x,y)$ is time translation
invariant and allows stationary eigenvectors of the form
$u_j(\vec{x})e^{i\omega x_4}$, which satisfies
\begin{equation}\label{spectral}
\int d^3y G^{-1}(\omega;\vec{x},\vec{y}) u_j(\vec{y})= i\gamma_4
\lambda_j(\omega) u_j(\vec{x}) \, ,
\end{equation}
where $G^{-1}(\omega;\vec{x},\vec{y})$ is the Fourier
transformation of $G^{-1}(x_4-y_4;\vec{x},\vec{y})$. The
eigenvalues $\lambda_j(\omega)$ can be written as the form
$\lambda_j(\omega)=\omega-i\epsilon_j(\omega)$, where $\epsilon_j$
is the quark eigen-energy of the state $j$. Calculating the
functional in Eq.(\ref{Eq}), one can know that the total energy is
the sum of the positive eigen-energy $\epsilon_j$, which satisfies
$\lambda(\epsilon_j) = 0 $, up to the chemical potential $\mu$.
These positive energy states satisfy the Dirac-like equation,
which can be given in momentum space as
\begin{equation}\label{quarkEq}
[i\gamma\cdot pA(p)+B(p)]u_j(\vec{p})+\frac{1}{f_\pi}\int
\frac{d^3k}{(2\pi)^3} B(\frac{p+k}{2}) [
\hat\sigma(\vec{p}-\vec{k}) + i\gamma_5 \vec\tau\cdot
\vec\pi(\vec{p}-\vec{k}) ] u_j(\vec{k})=0 \, .
\end{equation}
where $\hat{\sigma} = \sigma - f_\pi$. Because of the interaction,
there is a renormalization constant for the quark field, which can
be derived from the residue at the pole of Green's function and
given as\cite{GCMsoliton1}
\begin{equation}
Z_j=-\int d^3p d^3q \bar{u}_j(\vec{p}) \frac{\partial
G^{-1}(i\epsilon_j;\vec{p},\vec{q})}{\partial \epsilon_j}
u_j(\vec{q})  \,  .
\end{equation}
The total soliton energy functional of the soliton is then
\begin{equation}\label{solitonEnergy}
E=\sum\limits_{j=1}^3 \epsilon_j+\int d^3x \left[
\frac{1}{2}(\nabla \sigma)^2+\frac{1}{2}(\nabla \vec\pi)^2+U(\chi)
\right]  \, .
\end{equation}
The static meson fields are the configuration which minimizes the
energy functional, i.e., they are the solutions of the equations
$\delta E[\sigma,\vec\pi] /\delta \sigma = 0 $ and  $\delta
E[\sigma,\vec\pi] / \delta \vec\pi = 0 $. After some derivation,
one can write these equations explicitly as
\begin{eqnarray}
\label{sigmaEq} -\vec\nabla^2 \sigma(\vec{r})+\frac{\delta
U}{\delta \sigma(\vec{r})}+ Q_\sigma (\vec{r})  & = & 0 \, ,    \\
  \label{piEq}
-\vec\nabla^2 \vec{\pi}(\vec{r}) + \frac{\delta U}{\delta \vec\pi
 (\vec{r})} + Q_{\vec\pi} (\vec{r}) & = & 0 \, ,
\end{eqnarray}
where the $Q_\sigma$ and $Q_{\vec\pi}$ are the source terms
contributed from the valence quarks, and can be written as
\begin{eqnarray}
\label{quarksource_sigma} Q_\sigma(\vec{R}) & = &
\sum\limits_{j=1}^3 \frac{1}{f_\pi Z_j} \int d^3x d^3y
\bar{u}_j(\vec{x}) B(-\epsilon_j^2,\vec{x}-\vec{y})
\delta(\frac{\vec{x} + \vec{y}}{2}-\vec{R}) u_j(\vec{y}) \, ,   \\
 \label{quarksource_pi}
Q_{\vec\pi}(\vec{R}) & = & \sum\limits_{j=1}^3 \frac{1}{f_\pi Z_j}
\int d^3x d^3y \bar{u}_j(\vec{x}) B(-\epsilon_j^2,\vec{x}-\vec{y})
i\gamma_5 \vec{\tau} \delta(\frac{\vec{x}+\vec{y}}{2}-\vec{R})
u_j(\vec{y})     \, .
\end{eqnarray}

The above Eqs.(\ref{quarkEq}),
(\ref{solitonEnergy})-(\ref{quarksource_pi}) can be solved
self-consistently with numerical technique. When one wants to
solve these equations with the chiral circle constraint, one needs
only to discard the $\pi$ degrees of freedom or set $\pi$ as zero
to solve the other equations. And it has been carried out in
Refs.\cite{GCMsoliton1,GCMsoliton2}. In this work, we intend to
solve the equations beyond the chiral circle constraint.

\section{Numerical Calculation and Results}

To solve the soliton equations, one needs at first the effective
gluon propagator $D(x-y)$ to get the scalar functions $A(p^2)$ and
$B(p^2)$ and fix the quark propagator. In the present work, we
take the effective gluon propagator in the form of the
Munczek-Nominovsky model\cite{MN83}
\begin{equation} \label{Munm}
  g^2 D(q) =(2\pi)^4 \frac{3}{16} \eta^2 \delta^4(q) \, .
\end{equation}
Then the Dyson-Schwinger equation (Eq.(\ref{DSE})) yields
solutions
\begin{eqnarray}\label{AB}
A(p^2) & = &
    \left\{ \begin{array}{cc} 2 \, , & \quad p^2\leq \frac{\eta^2}{4} \, , \\
       \frac{1}{2}(1+ \sqrt{1+\frac{2\eta^2}{p^2}}) \, ,
   & \quad p^2 >  \frac{\eta^2}{4} \, ,
              \end{array}  \right.  \\
B(p^2) & = &
     \left\{ \begin{array}{cc} \sqrt{\eta^2-4p^2} \, ,
     & \qquad \quad p^2\leq \frac{\eta^2}{4} \, , \\
   0 \, , & \qquad \quad p^2 > \frac{\eta^2}{4} \, .
              \end{array}  \right.
\end{eqnarray}
This long range interaction has been reported to have the
dynamical confinement property \cite{GCMsoliton4}. Although the
effective gluon propagator can take a more realistic form, we
would show that this simple model can give a good data to compare
with the experiment.

From Eqs.~(\ref{quarkEq}),
(\ref{solitonEnergy})-(\ref{quarksource_pi}), one knows that they
are coupled Dirac-like equation in momentum-space and
Klein-Gorden-like equations in coordinate-space. As mentioned
above, this set of coupled equations has been solved with the
chiral circle constraint, i.e., setting $\vec{\pi} \equiv 0 $, in
Refs.\cite{GCMsoliton1,GCMsoliton2}. In this work, we solve the
equations with and beyond the chiral circle constraint with the
same methods as that in Refs.\cite{GCMsoliton1,GCMsoliton2}.

For the equation of quark field, we solve it in momentum space.
For a nucleon, we consider only the states with orbital angular
momentum $L=0$, and express the wavefunction of a valence quark as
\begin{equation}
u_j(\vec{p})=\left( \begin{array}{c} f_j(p) \\
i\vec{\sigma}\cdot\hat{p}g_j(p)  \end{array} \right)  \, .
\end{equation}
With the numerical integration and differentiating technique, we
transform the Dirac-like equation to an algebraic eigenvalue
equation $H_{mn}(\epsilon_j)X_n=0$, where $X_{2n}=f_j(nh)$ and
$X_{2n+1}=g_j(nh)$ with $h$ being the integration step of the
momentum, $\epsilon_j$ is the eigenvalue. If there exists a
nontrivial solution of $X$, $\det(H)$ should be zero. We search
for the eigenvalue $\epsilon_j$ from zero to a definite value
until $\det(H(\epsilon_j))=0$ and get the eigenvalue $\epsilon$
and eigenvector $X$ with the singular vector decomposition
method(SVD)\cite{SVD}.

For the equations of the meson fields, we solve them in coordinate
space with Newton's functional iteration method. Since the left
hand side of Eqs.(\ref{sigmaEq}) and (\ref{piEq}) is a functional
of $\sigma$ and $\vec\pi$, we can refer it to $F[\sigma,\vec\pi]$.
The meson fields ($\sigma_{s} , \vec{\pi}_{s}$) are the solutions
of $F[\sigma_s,\vec\pi_s]=0$. To solve these equations, we expand
the $F[\sigma,\vec{\pi}]$ at an initial value $\sigma=\sigma_0$,
$\vec\pi=\vec\pi_0$ and obtain
\begin{eqnarray}
 F[\sigma,\vec\pi] &=& F[\sigma_0,\vec\pi_0]
 +\int d^4x \left. \frac{\delta F[\sigma,\vec\pi]}{\delta \sigma}
 \right|_{\sigma=\sigma_0,\vec\pi=\vec\pi_0} \delta\sigma(x)
 \nonumber  \\
 & & +\int d^4x \left. \frac{\delta F[\sigma,\vec\pi]}{\delta \vec\pi}
 \right|_{\sigma=\sigma_0,\vec\pi=\vec\pi_0} \delta\vec\pi(x)
 +\cdots \, .
\end{eqnarray}
Neglecting the higher order and setting $F[\sigma,\vec\pi]$ as
zero, we gain an iteration equation of the meson fields,
\begin{equation}
\left\{ \begin{array}{rcl}
\sigma_{n+1}(x) & = & \sigma_{n}(x)+\delta\sigma_n(x) \, ,\\
\vec\pi_{n+1}(x)& = & \vec\pi_{n}(x)+\delta\vec\pi_n(x) \, ,
   \end{array} \right.
\end{equation}
where $\sigma_n(x)$ and $\vec{\pi}_{n}(x)$ are the solutions after
$n$-times iteration. It should be mentioned that, with the chiral
circle constraint, i.e., $\vec{\pi} \equiv 0$, the coupled equations
can be solved with less effort. However, when one solves the
equations beyond the chiral circle constraint, there are four
fields: the scalar field $\sigma$ and the triplet $\pi$ fields. It
is then very difficult to solve these equations. To the convenience
of numerical calculation, we take the Hedgehog form solution
\cite{Hedghog}
\begin{eqnarray}\label{hedgehog}
  \sigma(\vec{R}) = \sigma (R) \, , &                   \\
  \pi_{i} (\vec{R}) = \hat{R}_i \pi (R)  &  (i=1,2,3) \, .
\end{eqnarray}
which is one of the solution of the coupled equations.

In the calculation, we take at first a set of trial meson fields
to solve the Dirac-like equation and get the eigen-energy and
wave-function of the quark and the renormalization constant. With
the source terms constructed from the quark field, we solve the
meson field equations to get the new meson fields. Inserting the
new meson fields in the quark equation and repeating the above
steps until the meson fields and quark wave-function converge to a
desired precision, we obtain the final solutions.

One can recognize that $f_\pi$ is in proportion to $\eta$ from
Eq.(\ref{fpi}) when one takes the effective propagator in the form
of delta function in Eq.(\ref{Munm}) and gets the function $A$ and
$B$ as Eq.(\ref{AB}). One can then infer that the
Eqs.(\ref{quarkEq}), (\ref{sigmaEq}) and (\ref{piEq}) can be
scaled with the dimensional constant $\eta$. All the results can
thus be scaled to dimensionless data.

\begin{figure}[!htb]
\centering
\includegraphics[scale=1.3,angle=0]{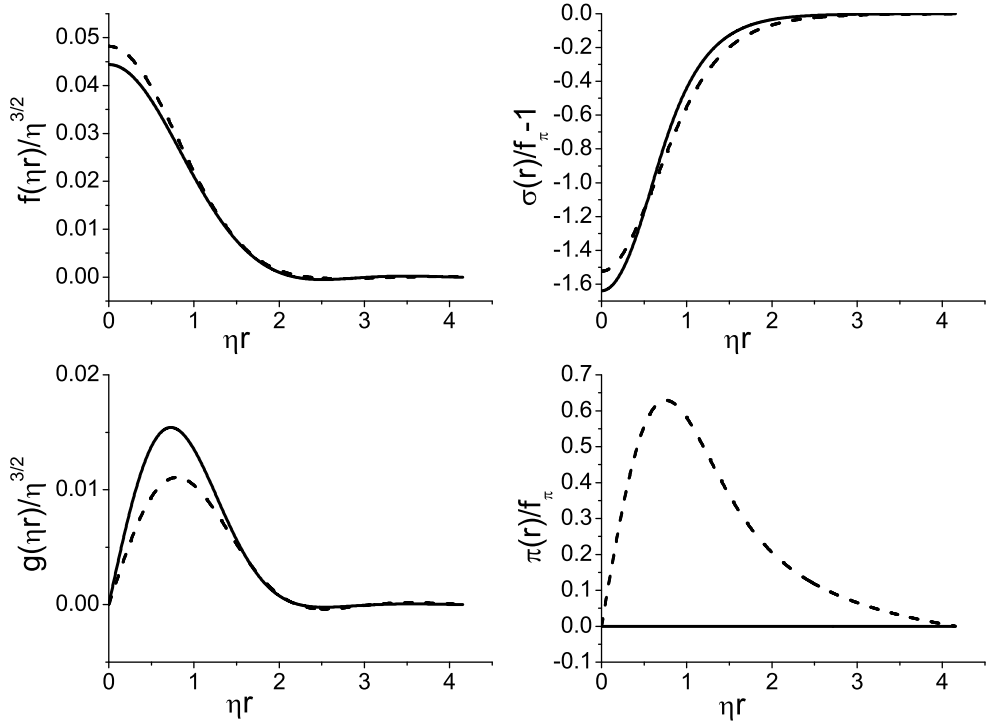}
\caption{\label{scaledResult} Rescaled quark and meson fields in
the soliton obtained in the numerical calculation. The thinner and
the thicker curves correspond to the results with and beyond
chiral circle constraint, respectively. The left top, bottom panel
shows the upper, lower component of the quark field accordingly.
The right top, bottom panel displays the $\sigma$, $\pi$ field
which is scaled with the pion coupling constant $f_\pi$,
respectively. }
\end{figure}

After solving the coupled equations, we obtain the eigen-energy
and wave-function of the quark and the meson fields (the obtained
results of the rescaled quark field and meson fields are
illustrated in Fig.~{\ref{scaledResult}} ), and further the
potential energy $E_{p}$ and kinetic energy $E_{k}$ of the meson
fields. The energy of the soliton can then be determined by $E= 3
\varepsilon_{j} + E_{p} + E_{k}$. Furthermore, We obtain the mass
of the soliton with two methods. One is the naive center mass
reduction, $M_s=\sqrt{E^2-3<p^2>}$, where $<p^2>$ is the
expectation value of the square of the quark momentum. The other
is the recoil correction\cite{recoilCorrection}.

\begin{figure}[htb]
\centering
\includegraphics[scale=1.3,angle=0]{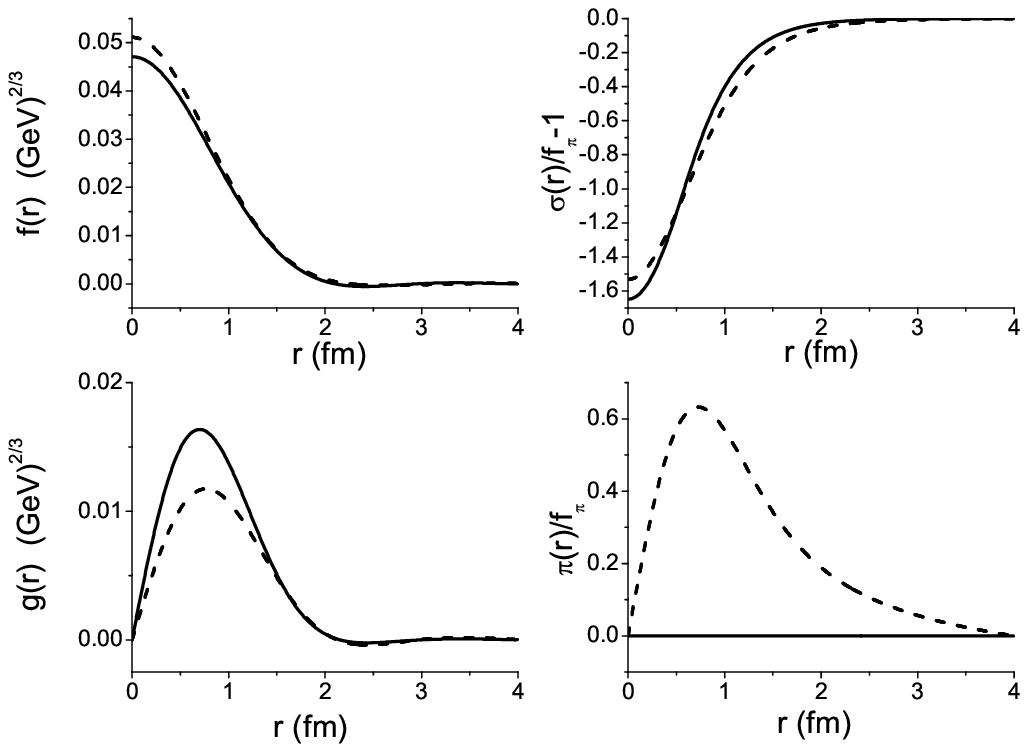}
\caption{\label{Graph1} Calculated quark and meson fields in the
soliton with and beyond chiral circle constraint in the case that
the energy scale $\eta=1.04$~GeV. The thinner, thicker curves
correspond to the results with and beyond chiral circle
constraint, respectively. The left top (bottom) panel shows the
upper (lower) component of the quark field. The right top (bottom)
panel displays the $\sigma$ ($\pi$) field which is scaled with the
pion coupling constant $f_\pi$. }
\end{figure}

In order to get a more impressive picture of the soliton, we adopt
explicitly the results with $\eta=1.04$~GeV, which is fixed by
fitting the experiment data $f_\pi=93$~MeV\cite{GCMsoliton2}. The
obtained properties of the soliton are listed in Table 1. The
obtained quark wave-function and the meson fields are illustrated
in Fig~\ref{Graph1}.

\begin{table}[ht]
\caption{\label{t1} Calculated mass and root mean square radius of
the soliton with and beyond the chiral circle constraint (with the
strength parameter $\eta=1.04$ in Eq.(\ref{Munm}) ) }
\vspace*{-6pt}
\begin{center}
\begin{tabular}{l|c|c}
\hline\hline
                                           & \quad with \qquad & beyond \\ \hline
 energy of valence quark $\epsilon$ (MeV)  &  339  &  129   \\
 meson potential energy(MeV)               &  218  &  109   \\
 meson kinetic energy(MeV)                 &  267  &  697   \\
 energy of soliton (MeV)                   & 1502  & 1193   \\
 mass (naive center of mass correction)(MeV)  & 1430  & 1124   \\
 mass (recoil correction)(MeV)             & 1020  &  916   \\
 radius (naive center of mass correction)(fm) & 0.71  &  0.64  \\
 radius (recoil correction)(fm)               & 0.59  &  0.56  \\ \hline\hline
\end{tabular}
\end{center}
\end{table}

From Table~\ref{t1} and the Fig~\ref{Graph1}, one can recognize
evidently that, besides the quite extended distribution, the pion
field provides a strong attractive interaction and reduces the
energy of the valence quark from $339$~MeV to $129$~MeV and
further the whole energy of the soliton from $1430$~MeV to
$1124$~MeV. From Fig~\ref{Graph1}, one can see that the $\sigma$
field beyond the chiral circle constraint is a little weaker than
that with the constraint. Then, the interaction between quarks
with $\sigma$ field is weaker. However, one can get a strong $\pi$
field beyond the chiral circle constraint, which provide a strong
attractive interaction between quarks. Thus, one can get a tighter
bound and lighter soliton beyond the chiral circle constraint.

In the calculation with chiral circle constraint, we have a state
with good quantum numbers, such as angular momentum and isospin.
Then we can compare the result with the experiment data of a
nucleon. The results listed in Table 1 show that the calculated
result does not agree with experimental data well. In the case of
beyond the chiral circle constraint, our calculation can not give
a state with good quantum numbers because of the inclusion of the
pion field and the Hedgehog approximation (The Hedgehog state is
invariant under the simultaneous Lorentz rotation and the isospin
rotation. In order to derive a state with good quantum numbers, we
should quantize the classical soliton). The state we obtained is
then a mixture of a nucleon and a delta. The result listed in
Table 1 indicates that the presently obtained mass of the soliton
with the naive correction on the center of mass is comparable to
the experimental data. However, in the case of the recoil
correction on the center of mass, the obtained mass is smaller
than the experimental data. Such a result is not strange since we
have not quantized the soliton to a physical state. Meanwhile, it
is promising to see that the soliton mass is close to the right.
When we quantize the soliton, the nucleon mass is about tens of
MeV heavier than the soliton mass. Then the mass of a nucleon is
nearly equal to the experiment data. The Table shows also that the
radius of the soliton is smaller than the nucleon charge radius
0.87~fm observed in experiment. It is quite natural because we
have taken the effective gluon propagator with infinite long range
interaction. The realistic interaction is surely not in infinite
long range. If we take a more realistic form with long range
decreasing behavior, we expect that a better fit to the experiment
data can be obtained.

\section{Summary and Remarks}

In this paper, we studied the GCM soliton with and beyond the
chiral circle constraint. The calculation shows that, when the
pion field is taken into account in the scheme beyond the chiral
circle constraint, the calculated mass of the GCM soliton is
comparable with experimental data. The calculation indicates that
the pion field provides a strong attraction, which binds the
quarks in the soliton more tightly and decreases the energy of the
soliton to a more realistic value.

In this calculation, we take the form of the effective gluon
propagator as a delta function in momentum space. In such a case,
the interaction between quarks is in infinite long range with a
constant strength. The calculated mean square root of the radius
of the soliton is then smaller than the experimental data. Since
the realistic form of the effective gluon propagator should
include the infrared enhancement, which is closely related to the
chiral symmetry breaking, and the ultraviolet property of
asymptotic freedom, which has a decrease on the interaction
strength, more precise calculation with a realistic effective
gluon propagator is needed to give a good description of the
experimental data of nucleon. Besides, the soliton we obtained
with the pion field being included explicitly does not possesses
the good quantum numbers of a nucleon. Then, a quantization on the
classical soliton and calculations of more physical observables
are necessary. The relevant works are under progress.

\bigskip

This work was supported by the National Natural Science Foundation
of China under contract Nos. 10425521, 10075002, 10135030, the
Major State Basic Research Development Program under contract No.
G2000077400 and the Research Fund for the Doctoral Program of
Higher Education of China under Grant No 20040001010. One of the
authors (Y.X. Liu) would also acknowledge the support of the
Foundation for University Key Teacher by the Ministry of
Education, China.


\newpage


\end{document}